\documentclass{PoS}
\usepackage{amssymb}
\usepackage{amsmath}   
\usepackage{times}
\usepackage{fontenc}
\usepackage{mathptmx}
\usepackage{graphicx}                                                     
\usepackage{slashed}
\usepackage{multirow}
\usepackage{epsfig,boxedminipage}

\title{Nucleon-Nucleon effective field theory with dibaryon fields}

\ShortTitle{Nucleon-Nucleon effective field theory with dibaryon fields}

\author{\speaker{Jaume Tarrús}
\\
        \small{\it{Departament d'Estructura i Constituents de la Mat\`eria and Institut de Ci\`encies del Cosmos}\\
        \small{\it{Universitat de Barcelona}}\\
        E-mail: \email{tarrus@ecm.ub.es}}}

\abstract{We report on a calculation of the nucleon-nucleon scattering amplitudes in the $^1S_0$ and $^3S_1$-$^3D_1$ channels at next-to-next to leading order using a recently proposed non-relativistic chiral effective theory, which includes dibaryon field as fundamental degrees of freedom.}

\FullConference{International Workshop on Effective Field Theories: from the pion to the upsilon \\
		February 2-6 2009\\
		Valencia, Spain}
\begin{document}


\section{Introduction}
\indent

In a recent paper \cite{Soto:2007pg} we proposed a chiral non-relativistic EFT which included two dibaryon fields \cite{Kaplan:1996nv} as a fundamental degrees of freedom. This EFT, which will be simply referred as NNEFT in this paper, is renormalizable and has simple counting rules when dimensional regularization (DR) and minimal subtraction (MS) scheme are used. The nucleon-nucleon scattering amplitudes in the $^1S_0$ and $^3S_1$ channels were calculated at NLO and a good description of data achieved in the $0-50 MeV$ energy range. We report here on a calculation at N$^2$LO, which was carried out in \cite{Soto:2009xy}, in order to see if the good description of data persists and check the convergence of the EFT. This is in fact mandatory in view of the fact that the so called KSW approach \cite{KSW,Kaplan:1998we} also produced a good description of data at NLO, but turned out to have a bad convergence in the $^3S_1$ at N$^2$LO \cite{Fleming:1999ee,Fleming:1999bs}.   
We will restrict ourselves to an energy range $E$ such that $E\ll m_\pi$, the pion mass, and $p=\sqrt{m_N E}\sim m_\pi$. Pion fields can then be integrated out leading to an EFT, which was already described in \cite{Soto:2007pg}, which we will call potential NNEFT (pNNEFT). For $p\sim m_\pi$ this EFT will already be suitable to carry out the calculations of the amplitudes. For $p\lesssim \frac{m^2_\pi}{\Lambda_\chi}$ however it will be convenient to integrated out nucleon fields with $p\sim m_\pi$ and use the so called pionless NNEFT ($\slashed{\pi}$NNEFT). All matching calculations will be done expanding the low energy or momentum scales in the integrals and using dimensional regularization to regulate any possible UV divergence and minimal subtraction scheme. Local field redefinitions which respect the counting will be used to get ride off redundant operators, rather than using the on-shell condition.   


\section{The nucleon-nucleon chiral effective theory with dibaryon fields} 

Our starting point is the effective theory for the $N_B$=2 sector of QCD for energies much smaller than $\Lambda_\chi$ (about $2m_N$, $m_N$ being the nucleon mass) recently proposed in \cite{Soto:2007pg}. The distinct feature of this EFT is that in addition to the usual degrees of freedom for a NNEFT theory, namely nucleons and pions, two dibaryon fields, an isovector ($D^a_s$) with quantum numbers $^1 S_0$ and an isoscalar ($\vec{D}_v$) with quantum numbers $^3 S_1$ are also included. Since $m_N \sim \Lambda_\chi$, a non-relativistic formulation of the nucleon fields is convenient \cite{JM}. Chiral symmetry, and its breaking due to the quark masses in QCD, constrain the possible interactions of the nucleons and dibaryon fields with the pions. The $N_B=0$ and $N_B=1$ sectors are the usual ones. We will need only the LO lagrangian in the $N_B=0$ sector and up to the NLO lagrangian in the $N_B=1$ sector.

The $N_B=2$ sector consist of terms with (local) two nucleon interactions, dibaryons and dibaryon-nucleon interactions. The terms with two nucleon interactions can be removed by local field redefinitions \cite{Beane:2000fi} and will not be further considered.  
The LO terms with dibaryon fields and no nucleons in the rest frame of the dibaryons read,

\begin{equation}
\mathcal{L}_{\mathcal{O}(p)}=\frac{1}{2} Tr\left[D_{s}^{\dag}\Bigl(-id_0+\delta_{m_s}'\Bigr)D_s\right]+
\vec{D}_v^{\dag}\Bigl(-i\partial_0+\delta_{m_v}'\Bigr)\vec{D}_v,
\label{dbLO}
\end{equation}
where $D_s=D^a_s\tau_a$ and $\delta_{m_i}'$, $i=s,v$ are the dibaryon residual masses, which must be much smaller than $\Lambda_\chi$, otherwise the dibaryon should have been integrated out as the remaining resonances have. The covariant derivative for the scalar (isovector) dibaryon field is defined as $d_0D_s=\partial_0D_s+\frac{1}{2}[[u,\partial_0u],D_s]$. The NLO pion-dibaryon lagrangian

\begin{equation}
\begin{split}
\mathcal{L}_{\mathcal{O}(p^2)}=
&s_1Tr[D_s(u\mathcal{M}^{\dag}u+u^{\dag}\mathcal{M}u^{\dag})D^{\dag}_s]+s_2Tr[D^{\dag}_s(u\mathcal{M}^{\dag}u+u^{\dag}\mathcal{M}u^{\dag})D_s]+\\
&+s_3Tr[D^{\dag}_sD_su_0u_0]+s_4Tr[D^{\dag}_sD_su_iu_i]+s_5Tr[D_s^{\dag}u_0D_su_0]+s_6Tr[D_s^{\dag}u_iD_su_i]+\\
&+v_1\vec{D}^{\dag}_v\cdot\vec{D}_vTr[u^{\dag}\mathcal{M}u^{\dag}+u\mathcal{M}^{\dag}u]+v_2\vec{D}^{\dag}_v\cdot\vec{D}_vTr[u_0u_0]+v_3\vec{D}^{\dag}_v\cdot\vec{D}_vTr[u_iu_i]+ \\ 
&+v_4(D^{i\dag}D^{j}+D^iD^{j\dag})Tr[u_iu_j],
\end{split}
\label{ordrep2}
\end{equation}
$s_i\sim 1/\Lambda_\chi$, $i=1,..,6$ and $v_j$, $j=1,..,4$ are low energy constants (LEC). The dibaryon-nucleon interactions will also be needed at NLO. They read
\begin{equation}
\begin{split}
\mathcal{L}_{DN}^{(LO)}=& \frac{A_s}{\sqrt{2}}(N^{\dag}\sigma^2\tau^a\tau^2N^*)D_{s,a}+
\frac{A_s}{\sqrt{2}}(N^{\top}\sigma^2\tau^2\tau^aN)D^{\dag}_{s,a}+\\
&+\frac{A_v}{\sqrt{2}}(N^{\dag}\tau^2\vec{\sigma}\sigma^2N^*)\cdot\vec{D}_v+\frac{A_v}{\sqrt{2}}(N^{\top}\tau^2\sigma^2\vec{\sigma} N)\cdot\vec{D}_v^{\dag},
\end{split}
\label{dn}
\end{equation}

\begin{equation}
\begin{split}
\mathcal{L}_{DN}^{(NLO)}=&\frac{B_s}{\sqrt{2}}(N^{\dag}\sigma^2\tau^a\tau^2\nabla^2N^*)D_{s,a}+\frac{B_s}{\sqrt{2}}(N^{\top}\sigma^2\tau^2\tau^a\nabla^2N)D^{\dag}_{s,a}+\frac{B_v}{\sqrt{2}}(N^{\dag}\tau^2\vec{\sigma}\sigma^2\nabla^2N^*)\cdot\vec{D}_v+\\
&+\frac{B_v}{\sqrt{2}}(N^{\top}\tau^2\sigma^2\vec{\sigma}\nabla^2 N)\cdot\vec{D}_v^{\dag}+\frac{B'_v}{\sqrt{2}} (\nabla_i N^{\dag}\tau^2\sigma^i\sigma^2\nabla_j N^*)D^j_v + \frac{B'_v}{\sqrt{2}}(\nabla_i N^{\top}\tau^2 \sigma^2 \sigma^i\nabla_j N) D_v^{j\dag},
\end{split}
\label{dn2}
\end{equation}
with $A_s,A_v\sim \Lambda_\chi^{-1/2}$, $B_s,B_v,B'_v\sim 1/\Lambda_\chi$.

\section{Dibaryon propagator and counting} 

The tree level dibaryon propagator expression $i/(-E+\delta_{m_i}'-i\eta)$ gets an important contribution to the self-energy due to the interaction with the nucleons as discussed in \cite{Soto:2007pg}

\begin{equation}
\frac{i}{-E+\delta_{m_i}'+i\frac{A_i^2m_Np}{\pi}} \qquad i=s,v,
\label{dbself}
\end{equation}
which is always parametrically larger than the energy $E$. The size of the residual mass can be extracted computing the LO amplitude using the propagator (\ref{dbself}) and matching the result to the effective range expansion,

\begin{equation}
\delta_{m_i}' \sim \frac{1}{\pi a_i}\lesssim \frac{m_{\pi}^2}{\Lambda_{\chi}} \qquad i=s,v.
\end{equation}
where $a_i$ are the scattering lengths of the $^1S_0$ and $^3S_1$ channels respectively. As a consequence the full propagator can be expanded. Moreover equation (\ref{dbself}) implies that the dibaryon field should not be integrated out unless $p\ll \delta_{m_i}'$, instead of $E\ll \delta_{m_i}'$ as the tree level expression suggests. Since $\delta_{m_i}' \lesssim \frac{m_{\pi}^2}{\Lambda_{\chi}}$, it should also be kept as an explicit degree of freedom in the so called $\slashed{\pi}$NNEFT. The LO expression for the dibaryon field propagator becomes for $p\sim m_{\pi}$ (pNNEFT), 

\begin{equation}
\frac{\pi }{A^2_i m_Np} \qquad i=s,v,
\label{dbexp}
\end{equation}
and for $p \lesssim \frac{m_{\pi}^2}{\Lambda_{\chi}}$ ($\slashed{\pi}$NNEFT),

\begin{equation}
\frac{i}{\delta_{m_i}'+i\frac{A_i^2m_Np}{\pi}} \qquad i=s,v.
\label{pilessprop}
\end{equation}
The expanded terms will be taken into account through an effective vertex, which we will denote with a cross in the dibaryon propagator. Higher order terms in this expansion will be equivalent to multiple insertions of this vertex. Furthermore for $p \gg \delta_{m_i}'$ the LO Lagrangian becomes both scale and $SU(4)$ (spin-flavor Wigner symmetric) invariant, if the interactions with pions are neglected \cite{Mehen:1999qs}. Note also that the pNNEFT propagator is suppressed  by a $m_{\pi}/\Lambda_{\chi}$ factor respect to the $\slashed{\pi}$NNEFT propagator.

Except for the above mentioned contributions to the self-energy of the dibaryon fields, which become LO, the calculation can be organized perturbatively in powers of $1/\Lambda_\chi$. Hence one expects that any UV divergence arising in higher order calculations will be absorbed in a low energy constant of a higher dimensional operator built out of nucleon, dibaryon and pion fields (note that the linear divergence in the self-energy can be absorbed in $\delta'_{m_i}$).
  
\section{Matching to pNNEFT}

For energies $E\sim m_\pi^2/\Lambda_\chi \ll m_\pi$, the pion fields can be integrated out. This integration produces nucleon-nucleon potentials and redefinitions of low energy constants. We will follow the strategy of \cite{Eiras:2001hu}, which was inspired in the formalism developed in \cite{Mont}.

\begin{figure}
\centerline{
\begin{tabular}{||cc|cc||} \hline \hline
\rule{0pt}{1.1cm}\raisebox{0.6cm}{(a)}&\includegraphics[height=1cm]{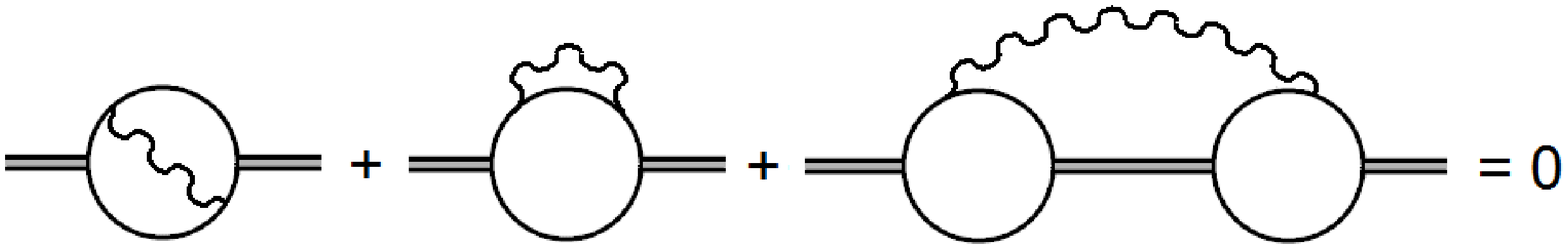}&\raisebox{0.6cm}{(b)}&\includegraphics[height=1cm]{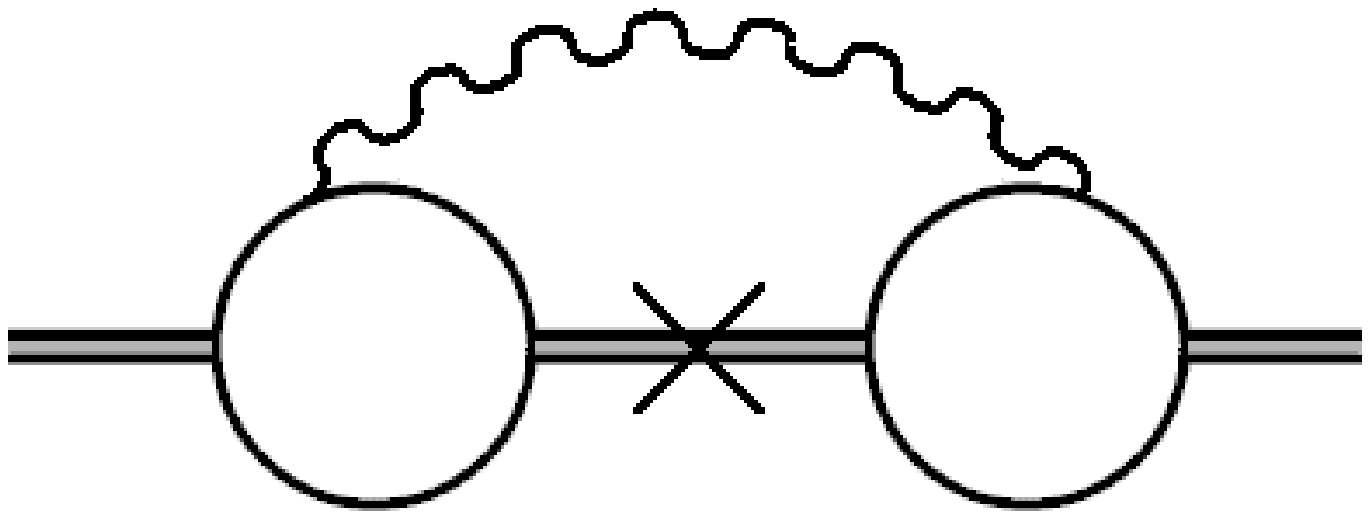} \\ \hline
\rule{0pt}{1.1cm}\raisebox{0.6cm}{(c)}&\includegraphics[height=1cm]{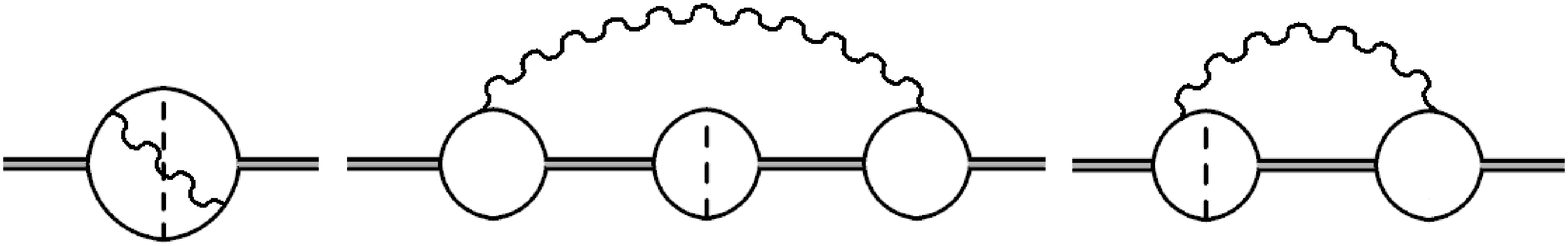}&\raisebox{0.6cm}{(d)}&\includegraphics[height=1cm]{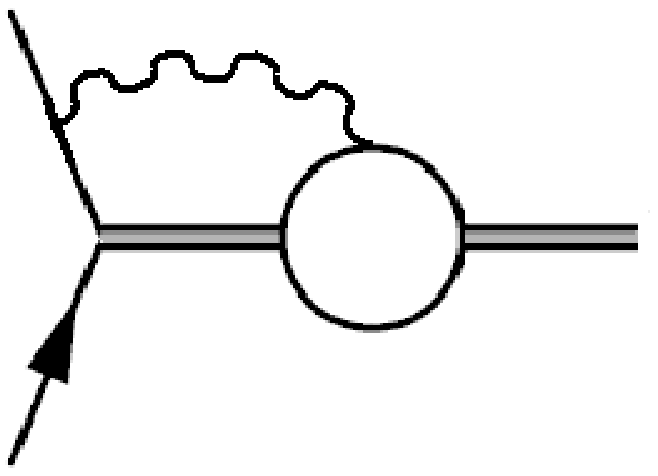}\\
\hline \hline
\end{tabular}}
\caption{\scriptsize{(a) and (b) are order $O(m_{\pi}^{3/2}/\Lambda_{\chi}^{3/2})$ contributions to the dibaryon residual mass. (a) These three diagrams sum up zero by Wigner symmetry. (b) Wigner symmetry is violated by insertions of $i(-E+\delta_{m_i})$. Naively we would expect these diagrams to be of higher order, $O(m_{\pi}^{5/2}/\Lambda_{\chi}^{5/2})$, but the energy term is enhanced by the radiation pion up to $O(m_{\pi}^2/\Lambda_{\chi}^2)$. Hence the cross in this diagram stands only for an insertion of the energy. (c) Order $O(m_{\pi}^2/\Lambda_{\chi}^2)$ contributions to the dibaryon residual mass, only the diagrams with the potential pion inside the radiation pion loop contribute. (d) N$^2$LO contribution to the dibaryon-nucleon vertex.}}
\label{matchinpnneff}
\end{figure}

The dibaryon residual masses get contributions from (\ref{ordrep2}) and higher loop diagrams involving radiation pions, like the ones in fig.\ref{matchinpnneff}b and fig.\ref{matchinpnneff}c,

\begin{equation}
\begin{split}
&\delta_{m_s}=\delta_{m_s}'+4m_q(s_1+s_2)+4A^2_s\frac{5}{3}\Bigl(\frac{g_A^2}{2f_{\pi}^2}\Bigr)^2 \Bigl(\frac{m_N m_{\pi}}{4\pi}\Bigr)^3+\Bigl(\frac{g_A^2}{2f_{\pi}^2}\Bigr)\frac{m^3_{\pi}}{8\pi},\\
&\delta_{m_v}=\delta_{m_v}'+4m_q v_1+4A^2_v\frac{5}{3}\Bigl(\frac{g_A^2}{2f_{\pi}^2}\Bigr)^2 \Bigl(\frac{m_N m_{\pi}}{4\pi}\Bigr)^3+\Bigl(\frac{g_A^2}{2f_{\pi}^2}\Bigr)\frac{m^3_{\pi}}{8\pi}.
\end{split}
\end{equation}
Note that because of $\delta_{m_i}'\lesssim \frac{m^2_\pi}{\Lambda_\chi}$ the quark mass dependence of $\delta_{m_i}$ is a leading order effect.

The dibaryon-nucleon vertices may in principle get $O(m_\pi^2/\Lambda_\chi^2)$ from a pion loop, but they turn out to vanish, except for those which reduce to iterations of the OPE potential which will already be included in the calculations in pNNEFT and must not be considered in the matching. There is however a two loop contribution of this order involving radiation pions from the diagram in fig.\ref{matchinpnneff}d.

\begin{equation}
A_i\rightarrow A_i \biggl(1-4\frac{g^2_A m^2_{\pi}}{(4\pi f_{\pi})^2}\biggr).
\end{equation}
In the $N_B=1$ sector the nucleon mass is redefined by a factor $\delta m_N$, however it can be reshuffled into $\delta_{m_i}$ by local field redefinitions. The dibaryon-nucleon interactions remain the same as in (\ref{dn}), except for the values of the $A_i$ which get modified.

Finally, in the $N_B=2$ sector, two nucleon non-local interactions (potentials) due to the one pion exchange are introduced. This is the well known one pion exchange (OPE) potential.

\begin{equation}
\begin{split}
\mathcal{L}_{N N}=& 
\frac{1}{2}\int d^3{\bf r} N^{\dag}\sigma^\alpha\tau^\rho N(x_1)V_{\alpha\beta ;\rho \sigma}(x_1-x_2)N^{\dag}\sigma^\beta\tau^\sigma N(x_2), 
\end{split}
\label{pot}
\end{equation}
with, 

\begin{equation}
V_{\alpha\beta,\rho\sigma}(x_1-x_2)=-\frac{g^2_A}{2f^2_{\pi}}\int\frac{d^3q}{(2\pi)^3} \frac{q_\alpha q_\beta}{\vec{q}^2+m^2_{\pi}} \delta^{\rho\sigma}e^{-i\vec{q}\cdot(\vec{x}_1-\vec{x}_2)}.
\label{ope}
\end{equation}

\section{Matching to $\slashed{\pi}$NNEFT}

For $p\lesssim \frac{m^2_\pi}{\Lambda_{\chi}}$ the calculation must be organized in a different way. This is very much facilitated if we integrate out nucleon three momenta of the order of $m_\pi$ first, which leads to the so called pionless nucleon-nucleon EFT. The Lagrangian of the $N_B=1$ sector of this theory remains the same as in pNNEFT. For the $N_B=2$ sector the only formal difference from pNNEFT is that the non-local potentials ({\ref{pot}) become local and can be organized in powers of $p^2/m_\pi^2$. Diagrams in Fig.\ref{schannel} containing one (or two) potential pion inside a nucleon bubble will contribute to the dibaryon time derivative term as well as the dibaryon residual mass. Contributions to the dibaryon time derivative can be reabsorbed by field redefinitions of dibaryon fields, while contributions to the residual mass simply redefine it. The derivative and non-derivative dibaryon-nucleon vertex get contributions from diagrams containing one (or two) potential pion in the dibaryon-nucleon vertex, redefining the LEC $B_i$ and $A_i$. Analogous diagrams involving three potential pion, not shown in this paper, will also analogously contribute to the matching. The OPE potential in (\ref{ope}) becomes $O(p^2/m_\pi^2\Lambda_\chi^2)$ and hence beyond N$^3$LO in the $p\ll m_{\pi}$ region.

\begin{figure}
\centerline{
\begin{tabular}{||cc|cc||} \hline \hline
\rule{0pt}{0.9cm}\raisebox{0.6cm}{LO}&\includegraphics*[height=0.8cm]{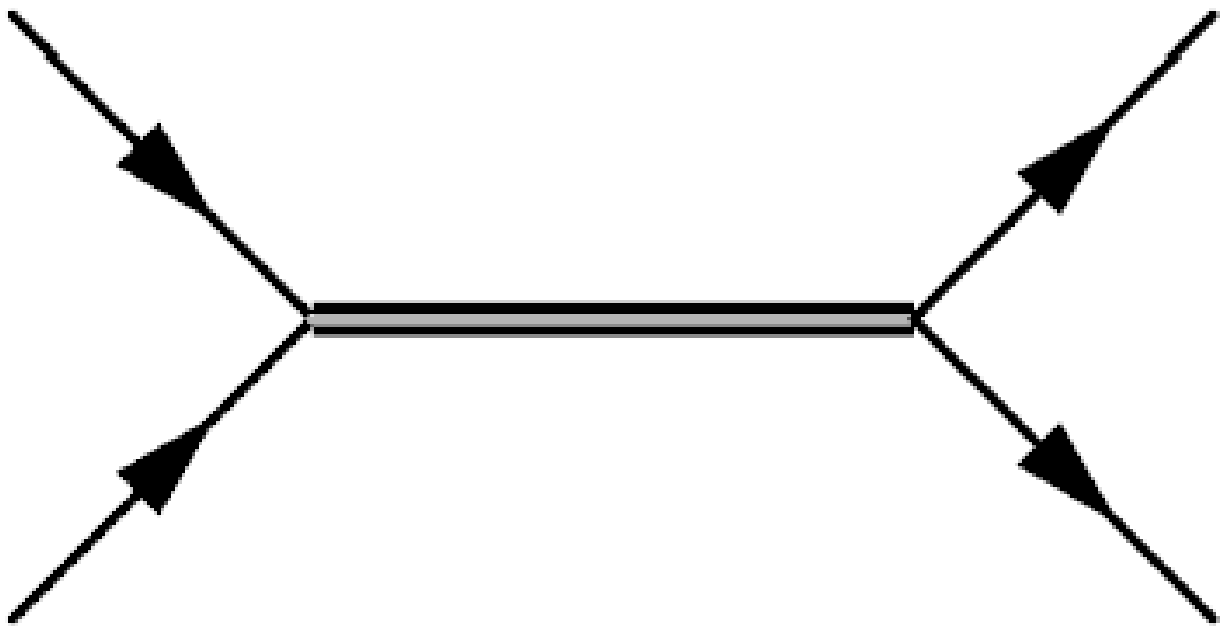}&
\raisebox{0.5cm}{N$^2$LO}&\multirow{2}{*}{\raisebox{0pt}[1.2cm]{\includegraphics*[height=1.7cm]{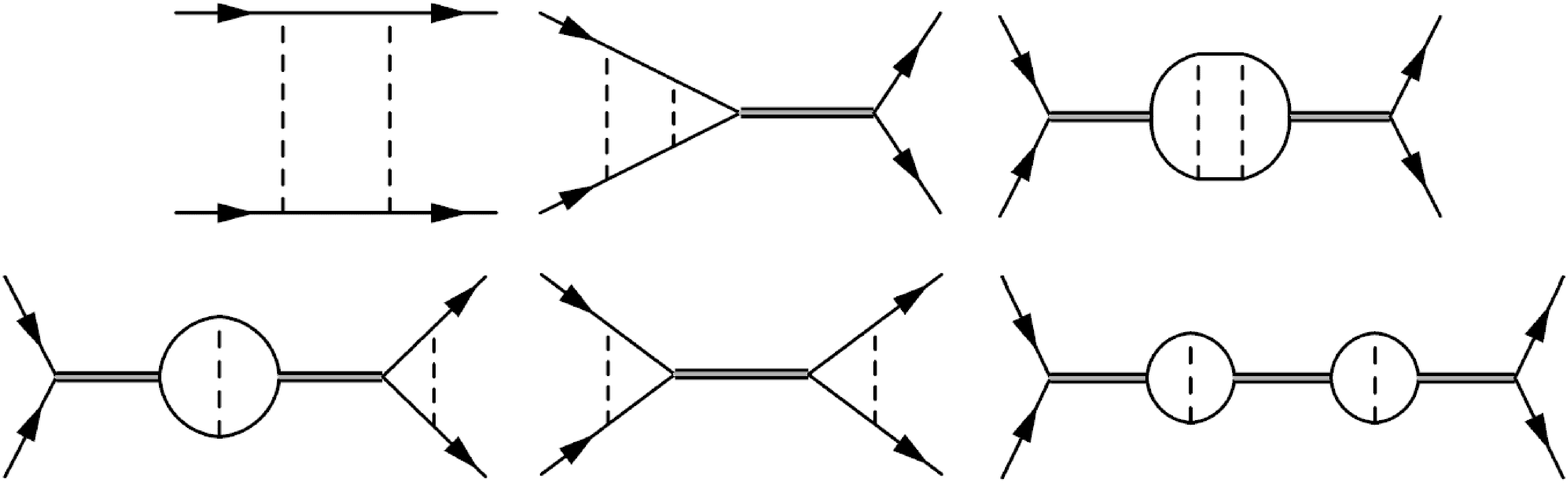}}}\\
\cline{1-2}
\rule{0pt}{0.9cm}\raisebox{0.5cm}{NLO}& \includegraphics*[height=0.8cm]{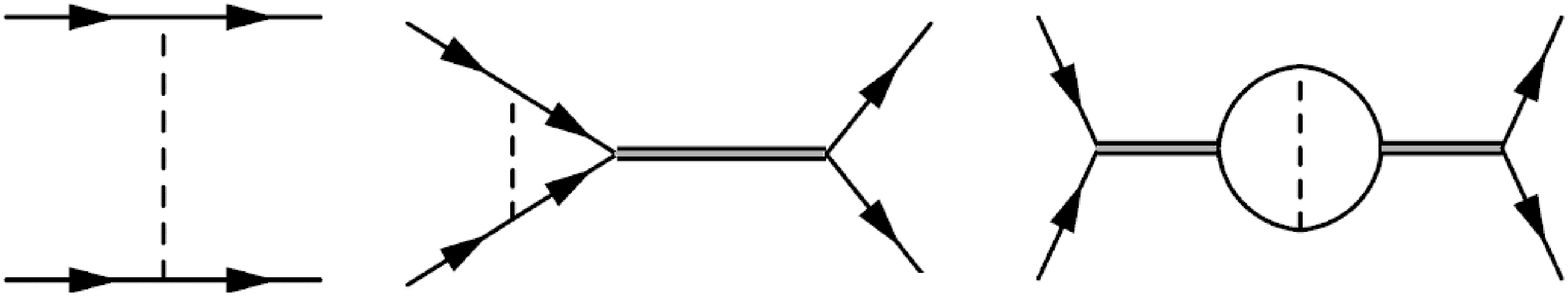} & &\\
\rule{0pt}{0.8cm}&\includegraphics*[height=0.8cm]{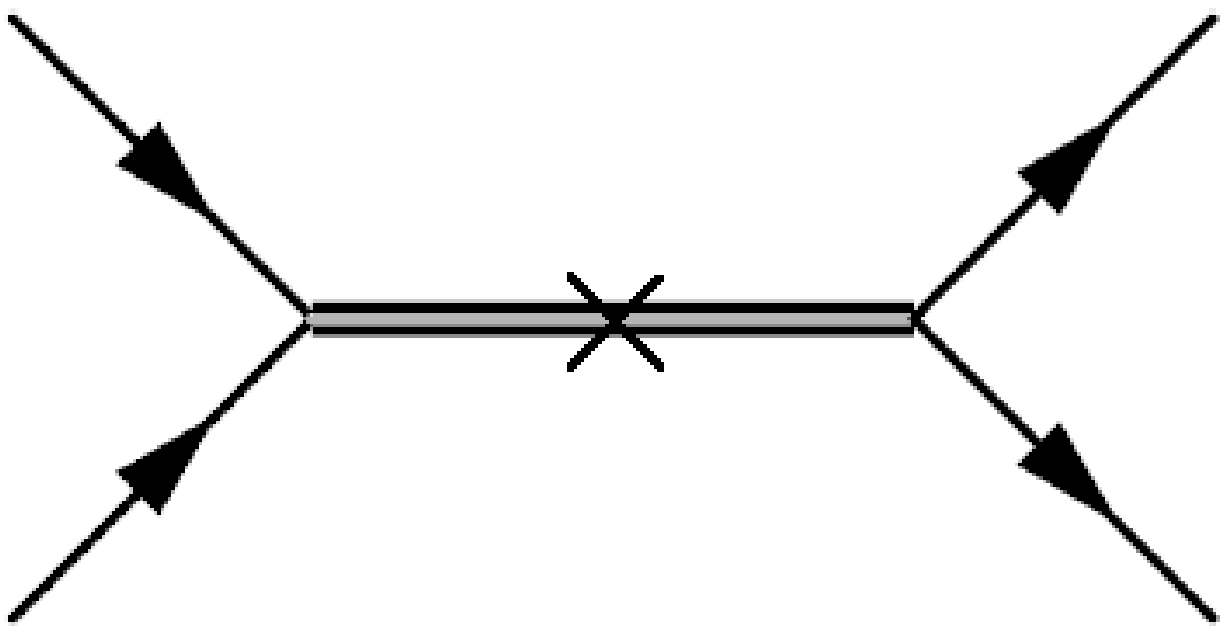} & &
\includegraphics*[height=0.8cm]{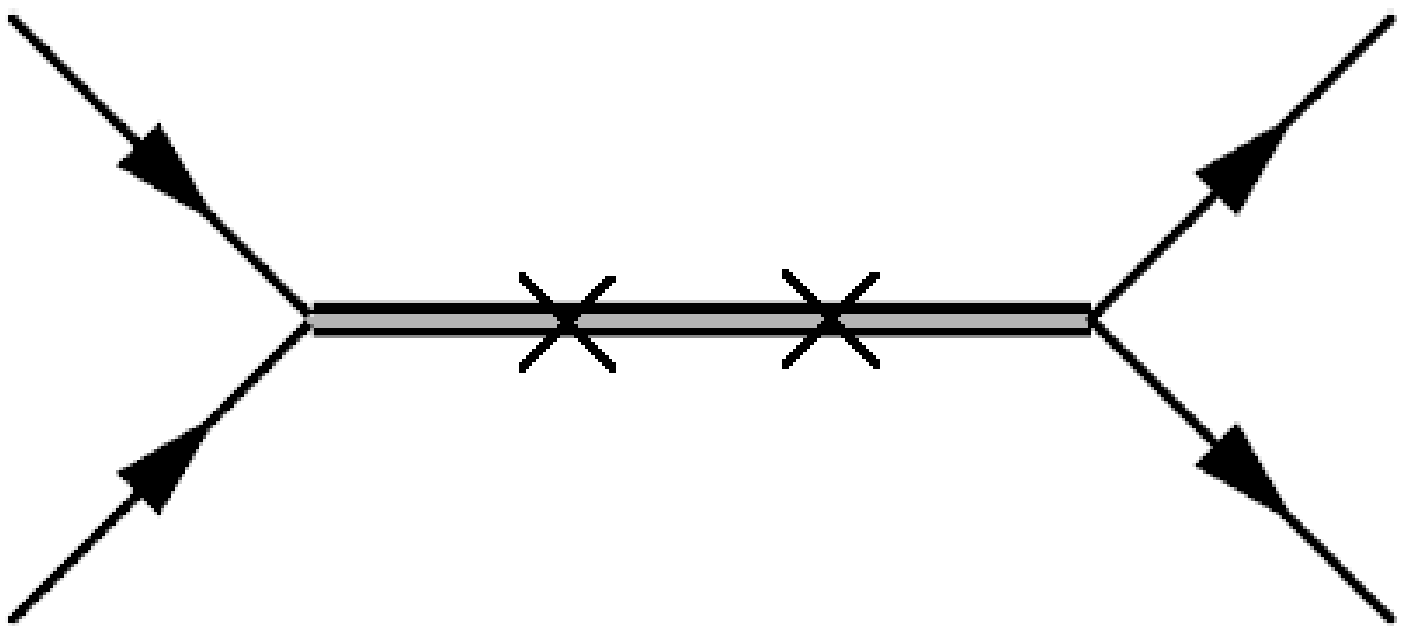}\includegraphics*[height=0.8cm]{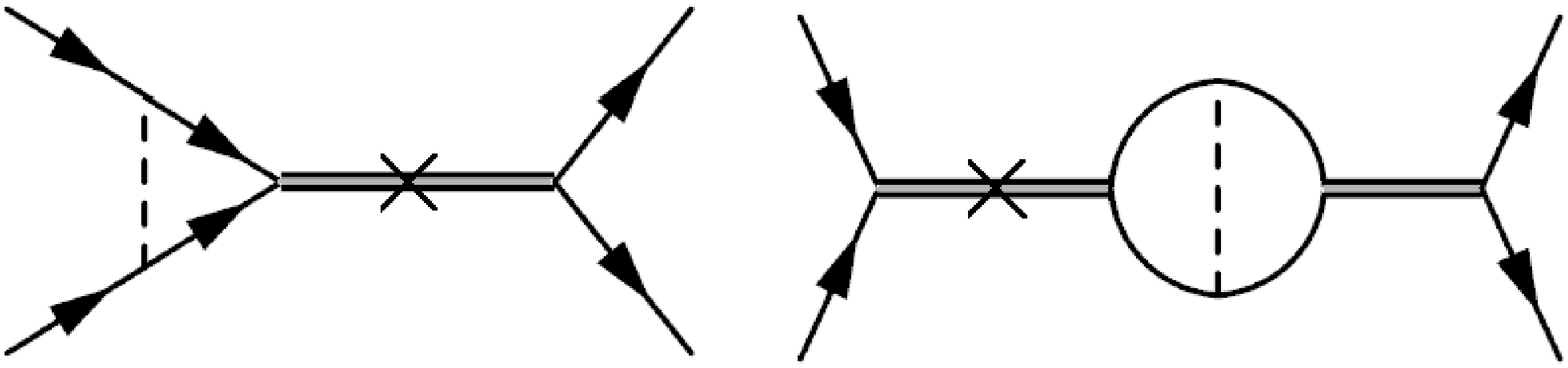} \\
\hline \hline
\end{tabular}}
\caption{Diagrams contributing to the $^1S_0$ and $^3S_1$ partial wave amplitudes.}
\label{schannel}
\end{figure}

\begin{figure}
\begin{minipage}[b]{0.5\linewidth}
\centerline{
\begin{tabular}{||cc||}
\hline \hline
\rule{0pt}{0.9cm}\raisebox{0.5cm}{NLO}&\includegraphics[height=0.8cm]{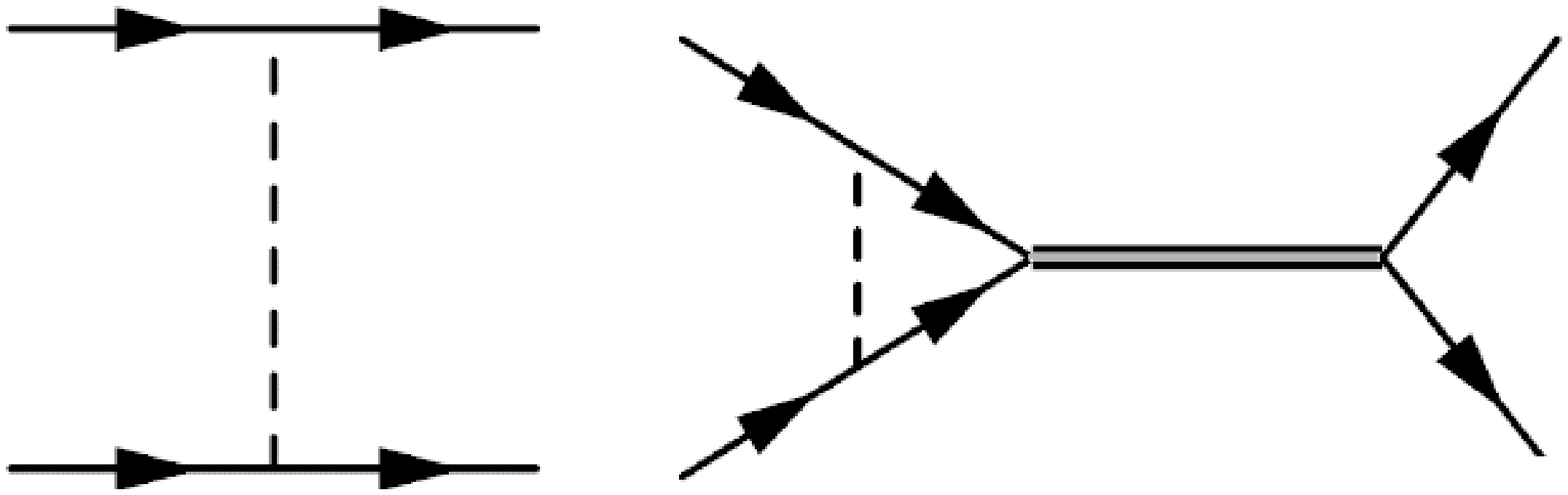}\\ \hline
\rule{0pt}{2.5cm}\raisebox{2.1cm}{N$^2$LO}&\includegraphics[height=2.4cm]{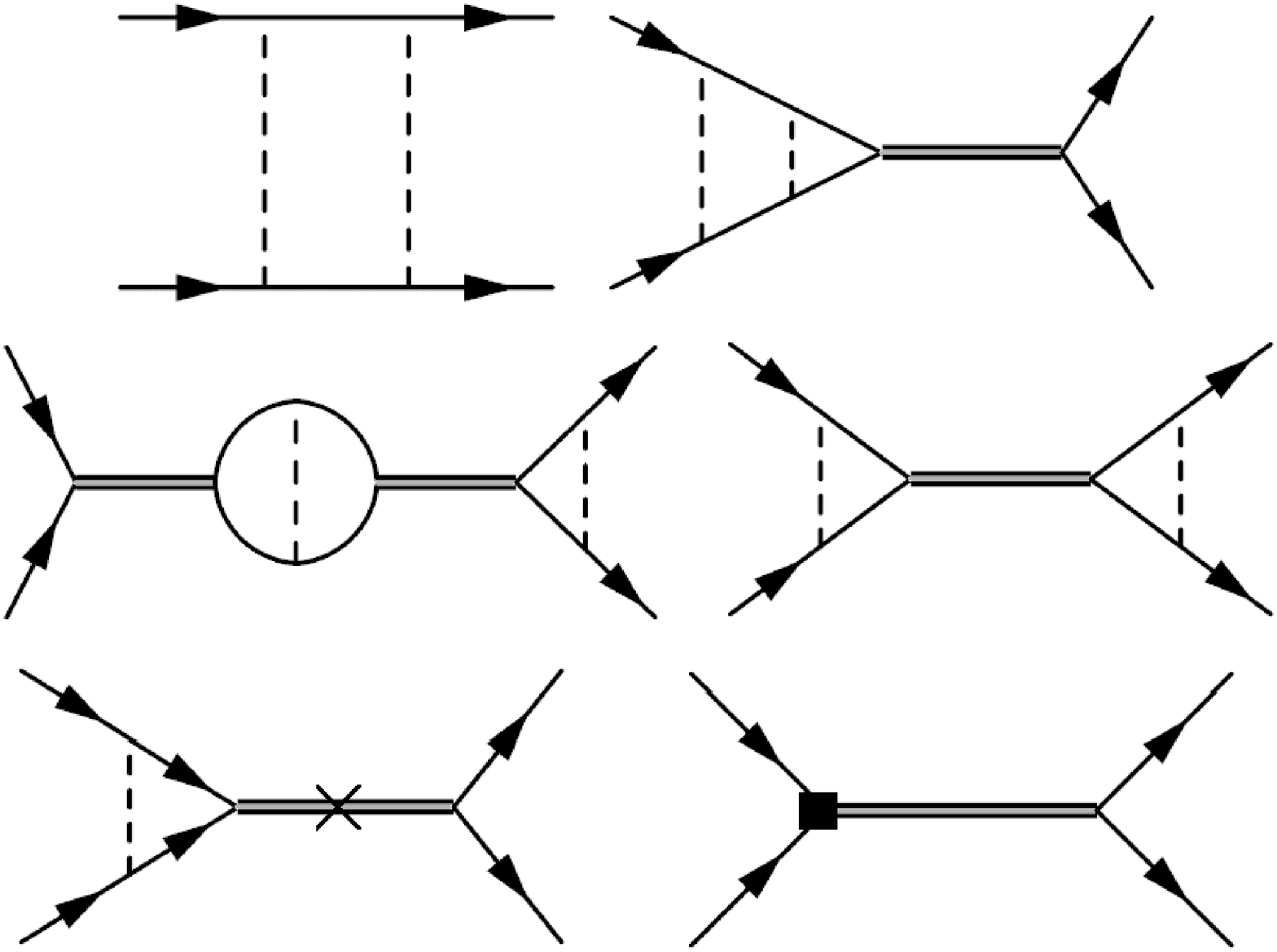}\\ \hline \hline
\end{tabular}}
\caption{Diagrams contributing to the mixing amplitude.}
\label{mixingdiag}
\end{minipage}
\hspace{0cm}
\begin{minipage}[b]{0.5\linewidth}
\centerline{
\begin{tabular}{||cc||}
\hline \hline
\rule{0pt}{0.9cm}\raisebox{0.5cm}{NLO}&\includegraphics[height=0.8cm]{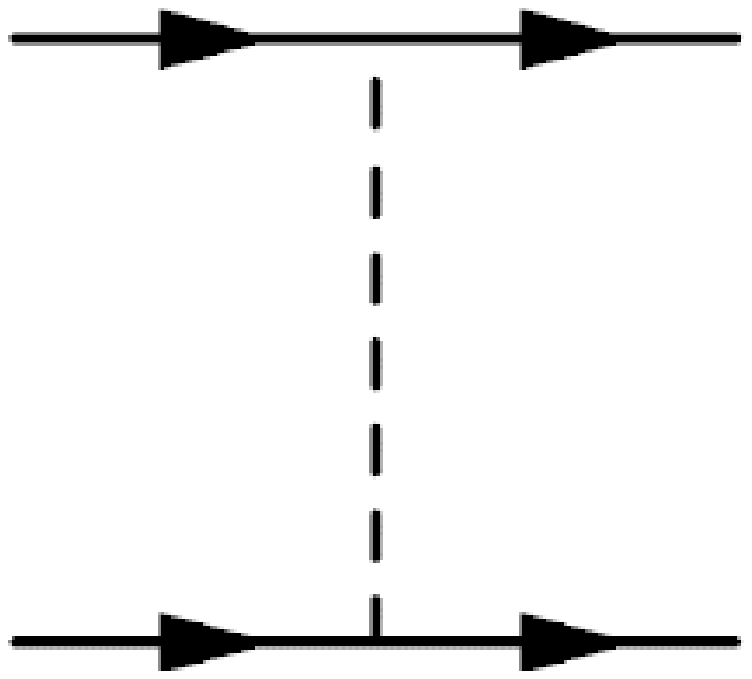}\\ \hline
\rule{0pt}{0.9cm}\raisebox{0.5cm}{N$^2$LO}&\includegraphics[height=0.8cm]{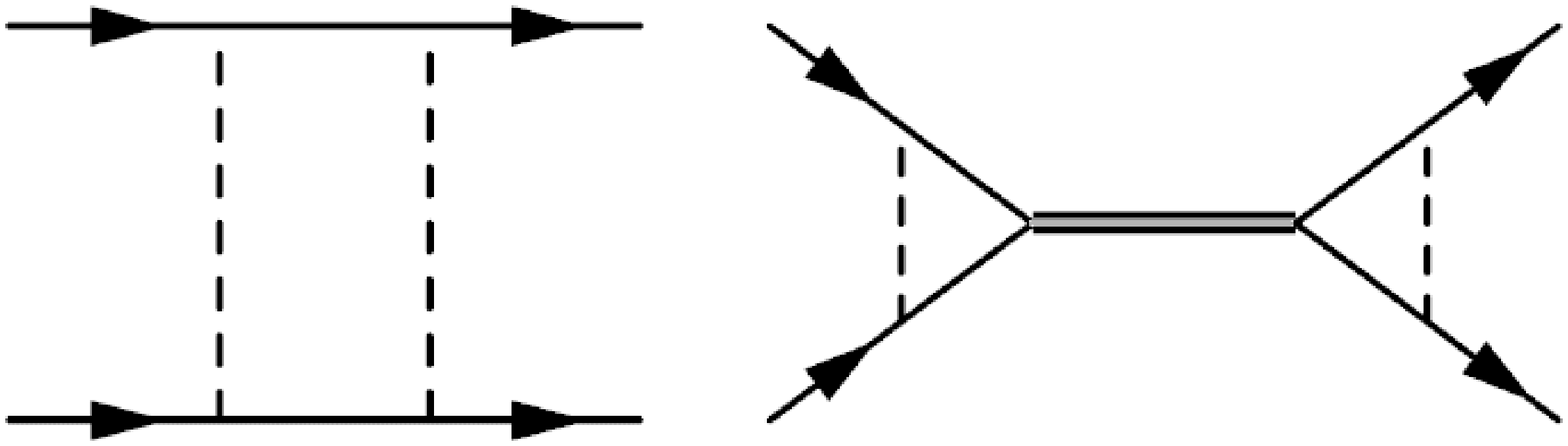}\\ \hline \hline
\end{tabular}}
\caption{Diagrams contributing to the $^3D_1$ partial wave amplitude.}
\label{3d1diag}
\end{minipage}
\end{figure}

\section{The $^1 S_0$ channel}

\paragraph{}The $^1S_0$ amplitude is computed from the diagrams in Fig.\ref{schannel} \cite{Soto:2009xy}. The phase shifts are plotted in Fig.\ref{1s0nlo} and Fig.\ref{1s0nnlo} and correspond in the 0-5MeV range to the $\slashed{\pi}$NNEFT and in the 5-50MeV range to pNNEFT. Parameters have been fit to phase shift data, in the 0-5MeV range in the case of the $\slashed{\pi}$NNEFT and to 5-20MeV range in the case of the pNNEFT. The reason behind the election to fit the pNNEFT to the 5-20MeV data range instead of the more natural 5-50MeV data range is that the former delivers a more smoother joining between the $\slashed{\pi}$NNEFT and the pNNEFT curves. Furthermore the convergence behavior is emphasized. Error bands correspond to the size of the next order in the phase shift series.  

Both $A_s$ and $\delta_{m_s}$ receive N$^2$LO corrections when matching from NNEFT to pNNEFT. If the whole expressions for $A_s$ and $\delta_{m_s}$ were to be used in the N$^2$LO amplitude, higher order terms will be introduced. Therefore we will differentiate between $\delta^{NLO}_{m_s}(A^{NLO}_s)$ and $\delta^{N^2LO}_{m_s}(A^{N^2LO}_s)$ and we will plug them in $\delta^{s,N^2LO}$ and $\delta^{s,NLO}$ respectively. The values of this parameters have been obtained by minimizing the sum of the $\chi^2$ functions associated to $\delta^{NLO}$ and $\delta^{N^2LO}$, with the errors given by the size of the next order in the phase shift series (i.e $(m_{\pi}/\Lambda_{\chi})^2$ and $(m_{\pi}/\Lambda_{\chi})^3$ respectively). 

The $\slashed{\pi}$NNEFT amplitude expressions for N$^2$LO (i.e. NLO in the pNNEFT counting) and N$^3$LO (i.e. N$^2$LO in the pNNEFT) are formally identical, however theoretical errors are smaller in the later. The only diagrams involved are the tree level one (LO) and the tree level with an insertion of a cross (NLO and N$^2$LO). The $\slashed{\pi}$NNEFT phase shift have been fitted independently of pNNEFT. Results for the $^1S_0$ channel parameters are summarized in Table \ref{taula1}.

\begin{table}
\begin{minipage}[b]{0.5\linewidth}
\scriptsize{\centerline{
\begin{tabular}{|c|c|c|c|}
 \cline{3-1} \cline{4-1}
 \multicolumn{2}{c|}{} & $A_s$ (Mev$^{-1}$) & $\delta_{m_s}$ (MeV) \\ \hline 
 \multicolumn{2}{|c|}{$p \lesssim m_\pi^2/\Lambda_\chi$} &  $0.0243$ & $-1.51$  \\ \hline
 \multirow{2}{*}{$p\sim m_{\pi}$} & NLO   & $0.0302$ & $-2.25$ \\ \cline{3-3} \cline{2-3} \cline{4-3}
  & N$^2$LO & $0.0251$ & $-10.5$ \\ 
   \hline
\end{tabular}}}
\caption{\scriptsize{Parameters for the $^1S_0$ channel.}}
\label{taula1}
\end{minipage}
\hspace{0cm}
\begin{minipage}[b]{0.5\linewidth}
\scriptsize{\centerline{
\begin{tabular}{|c|c|c|c|c|}
 \cline{3-1} \cline{4-1} \cline{4-5}
 \multicolumn{2}{c|}{} & $A_v$ (Mev$^{-1}$) & $\delta_{m_v}$ (MeV) & $B_v'/A_v$ (MeV$^{-2}$)\\ \hline 
 \multicolumn{2}{|c|}{$p \lesssim m_\pi^2/\Lambda_\chi$} &  $0.0265$ & $7.68$ & \\ \hline
 \multirow{2}{*}{$p\sim m_{\pi}$} & NLO   & $0.0359$ & $13.4$ & $12.4e-6$\\ \cline{3-3} \cline{2-3} \cline{4-3} \cline{4-5}
  & N$^2$LO & $0.0110$ & $45.3$ & \\ \cline{4-5}
   \hline
\end{tabular}}} 
\caption{\scriptsize{Parameters for the $^3S_1-^3D_1$ channel.}}
\label{taula2}
\end{minipage}
\end{table}

\begin{figure}
\begin{minipage}[b]{0.5\linewidth}
\centerline{
\includegraphics[width=4cm]{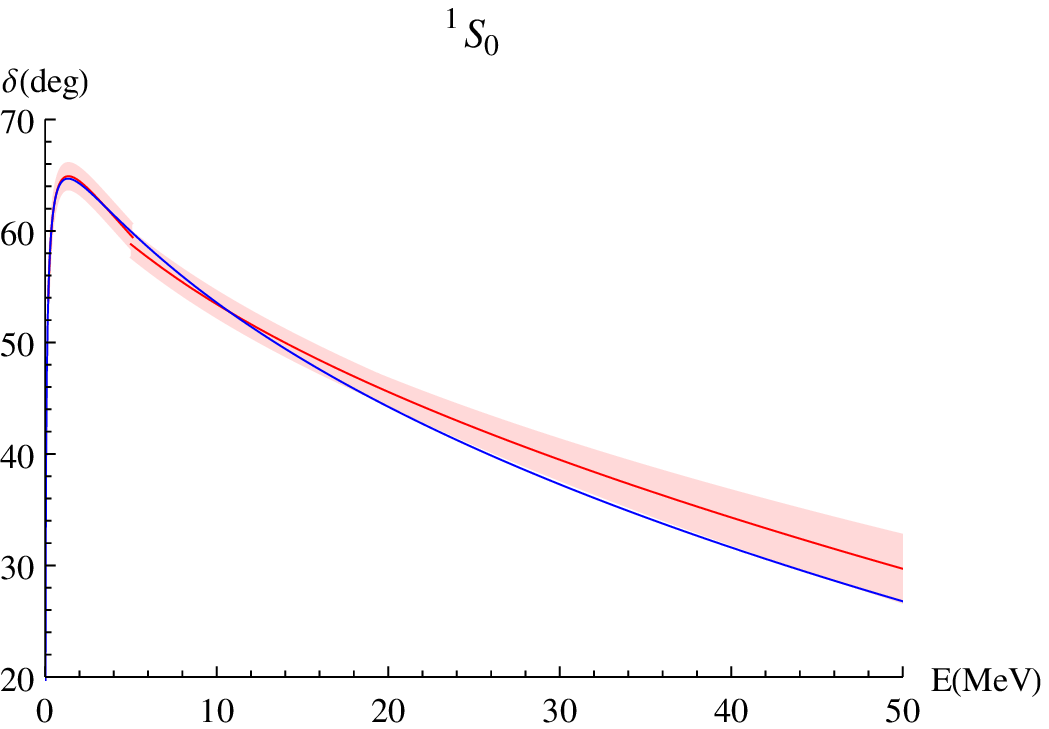}}
\caption{\scriptsize{Plot of the NLO expression for the $^1S_0$ phase shift. The blue line shows the Nijmegen data for the $^1S_0$ phase shift, while the red line corresponds to the fit of our expression.}}
\label{1s0nlo}
\end{minipage}
\hspace{1cm}
\begin{minipage}[b]{0.5\linewidth}
\centerline{\includegraphics[width=4cm]{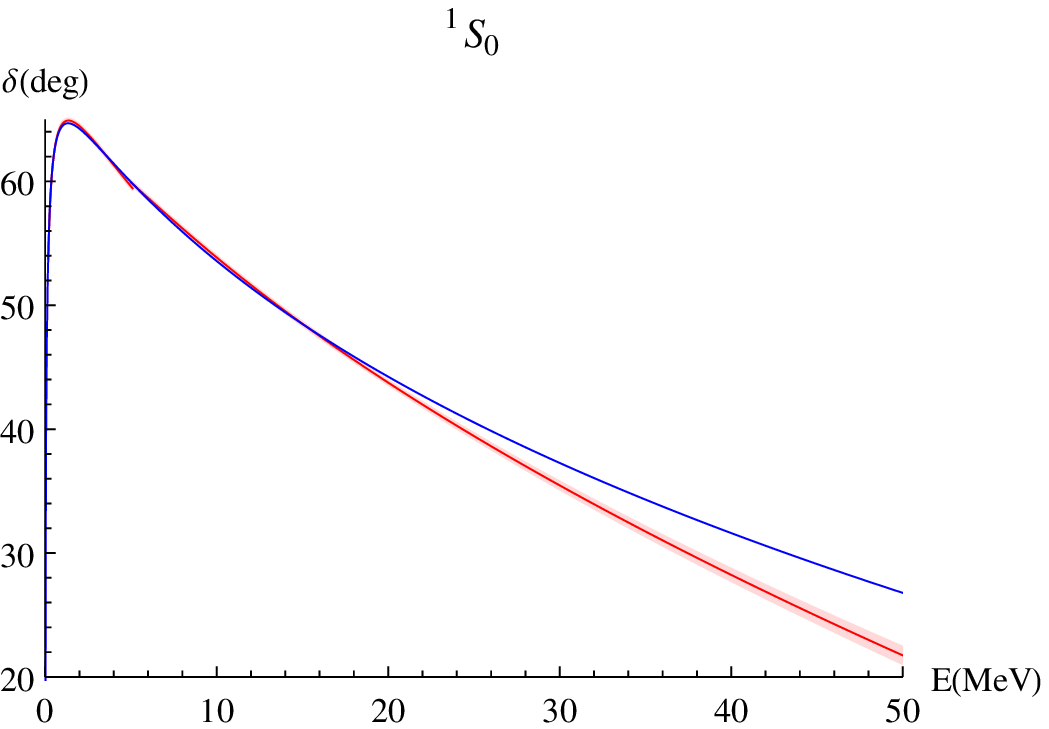}}
\caption{\scriptsize{Plot of the N$^2$LO expression for the $^1S_0$ phase shift. As in the previous figure the blue line shows the Nijmegen data for the $^1S_0$ phase shift.}}
\label{1s0nnlo}
\end{minipage}
\end{figure}

\section{The $^3 S_1$-$^3 D_1$ channel}

\paragraph{}The $^3S_1$ amplitude is computed from the diagrams in Fig.\ref{schannel} \cite{Soto:2009xy}, the mixing amplitude from the diagrams in Fig.\ref{mixingdiag}, and the $^3D_1$ amplitude from the diagrams in Fig.\ref{3d1diag}. The $^3S_1$ and $^3D_1$ phase shifts are plotted in Fig.\ref{3s1nlo}, Fig.\ref{3s1nnlo} and Fig.\ref{3d1} respectively. The mixing angle is plotted in Fig.\ref{mixing}. In this section we analyze $^3 S_1$-$^3 D_1$ channel. We compare the $^3 S_1$ and $^3 D_1$ phase shifts to data as well as the mixing angle. The fitting procedure is analogous to the one used for the $^1S_0$ channel. However in this channel we have minimized the sum of the $\chi^2$ functions associated to $\delta^{v,NLO}$ and $\epsilon^{N^2LO}$ obtaining the values of $A^{NLO}_v$,$\delta^{NLO}_{m_v}$ and $B'_v/A_v$. If we add the $\chi^2$ function associated to $\delta^{v,N^2LO}$ the minimization would not converge. Hence $A^{N^2LO}_v$ and $\delta^{N^2LO}_{m_v}$ have been obtained in a separated fit plugging in the values for the NLO parameters from the first fit. $\epsilon^{NLO}$, $\delta^{^3D_1,NLO}$ and, $\delta^{^3D_1,N^2LO}$ do not contain free parameters. The $\delta^{^3S_1}$ plotted correspond in the 0-5MeV range to the $\slashed{\pi}$NNEFT and in the 5-50MeV range to pNNEFT, the $\delta^{^3D_1}$ and $\epsilon$ plotted correspond to pNNEFT in the whole range. Results for the $^3S_1$-$^3D_1$ channel parameters are summarized in Table \ref{taula2}.

\begin{figure}
\begin{minipage}[b]{0.5\linewidth}
\centerline{
\includegraphics[width=4cm]{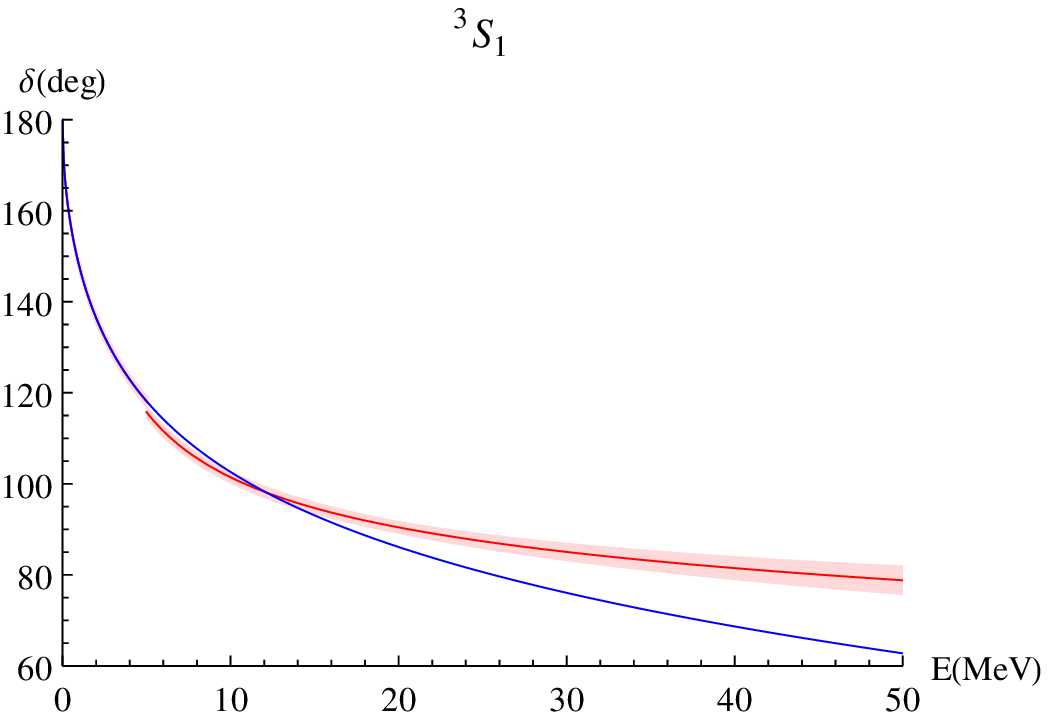}}
\caption{\scriptsize{Plot of the NLO expression for the $^3S_1$ phase shift. The blue line shows the Nijmegen data for the $^3S_1$ phase shift, while the red line corresponds to the fit of our expression.}}
\label{3s1nlo}
\end{minipage}
\hspace{1cm}
\begin{minipage}[b]{0.5\linewidth}
\centerline{\includegraphics[width=4cm]{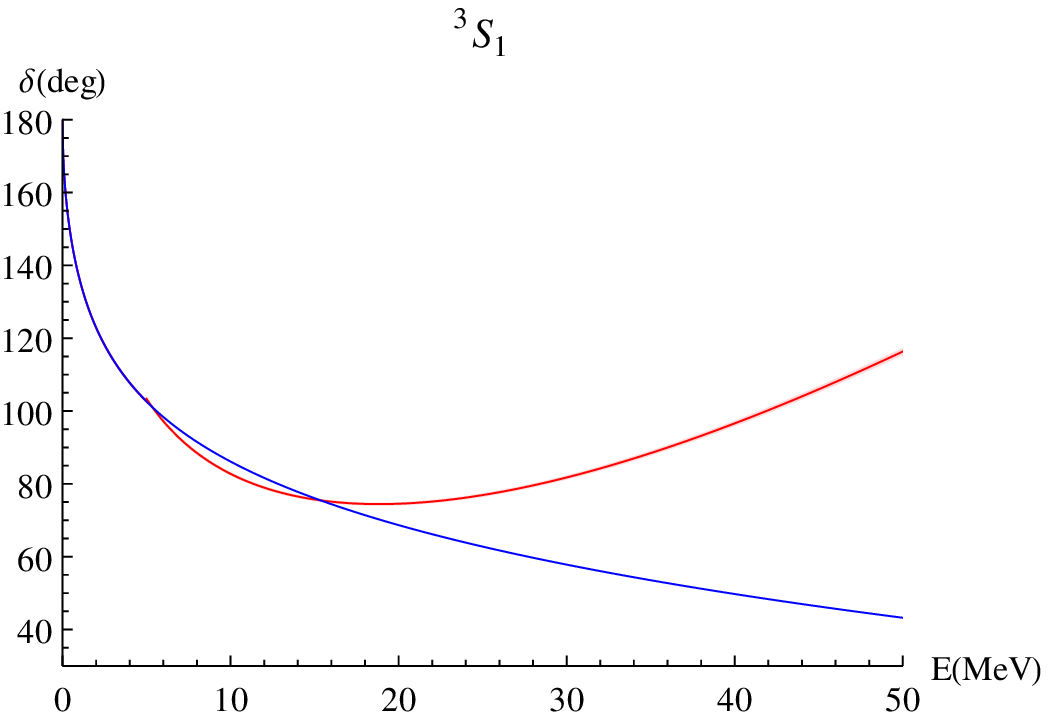}}
\caption{\scriptsize{Plot of the N$^2$LO expression for the $^3S_1$ phase shift. The blue curve is the Nijmegen data for the $^3S_1$ phase shift, while red line corresponds to our N$^2$LO expression.}}
\label{3s1nnlo}
\end{minipage}
\end{figure}

\begin{figure}
\begin{minipage}[b]{0.5\linewidth}
\centerline{
\includegraphics[width=4cm]{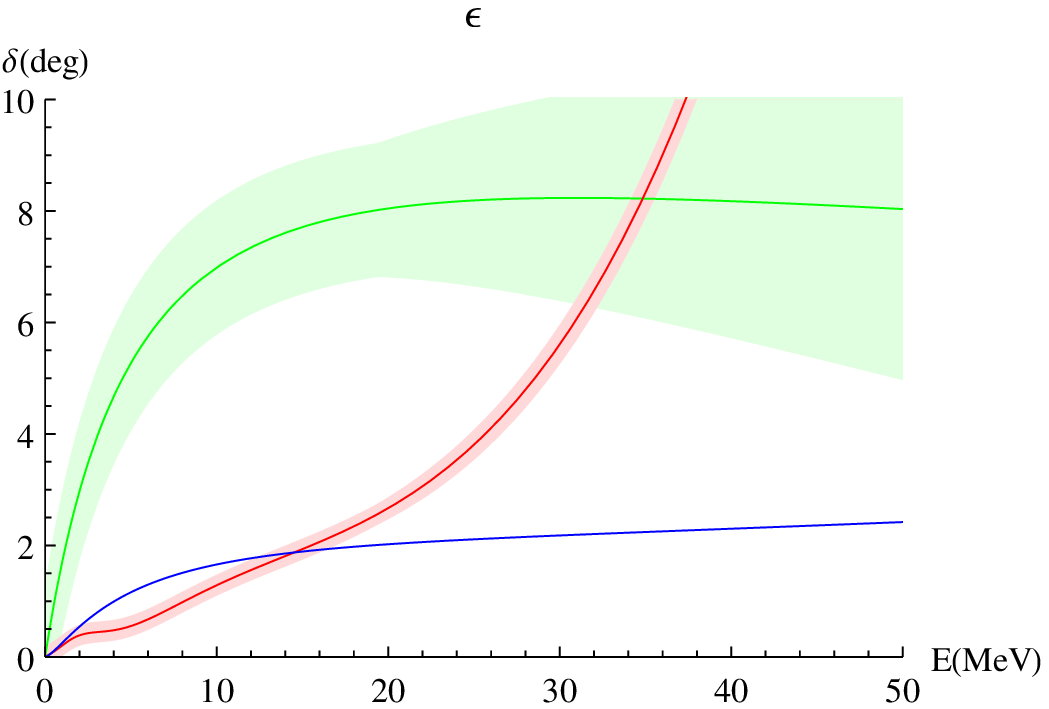}}
\caption{\scriptsize{Plot of the mixing angle. The blue line shows the Nijmegen data, the green and red lines the NLO and N$^2$LO expression respectively.}}
\label{mixing}
\end{minipage}
\hspace{1cm}
\begin{minipage}[b]{0.5\linewidth}
\centerline{\includegraphics[width=4cm]{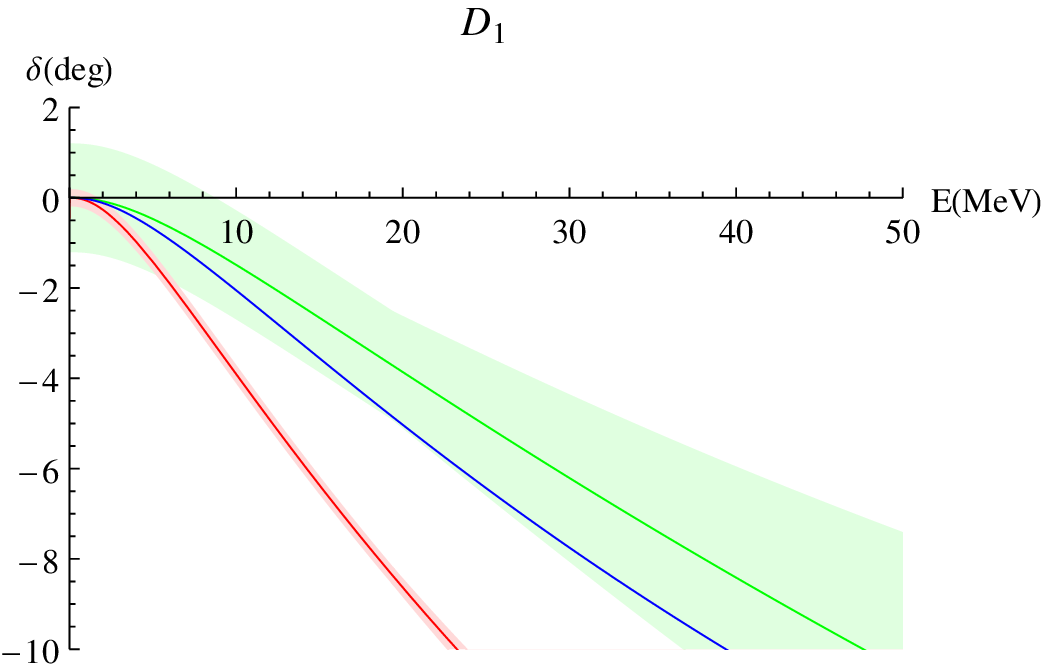}}
\caption{\scriptsize{Plot of the $^3D_1$ phase shift. The blue line shows the Nijmegen data, the green and red lines the NLO and N$^2$LO expression respectively.}}
\label{3d1}
\end{minipage}
\end{figure}

\section{Conclusions}

\paragraph{}We have calculated the nucleon-nucleon scattering amplitudes for energies smaller than the pion mass in the $^1S_0$ and the $^3S_1$-$^3D_1$ channels at N$^2$LO in a chiral effective field theory which contains dibaryon fields as fundamental degrees of freedom. The large scattering lengths in the $^1S_0$ and the $^3S_1$ channels force the dibaryon residual masses to be much smaller than the pion mass. We organize the calculation in a sequence of effective theories, which are obtained by sequentially integrating out higher energy and momentum scales. We first integrate out energy scales of the order of the pion mass. This leads to an effective theory with dibaryon and nucleon fields, pNNEFT. The latter interact through potentials. For three momenta of the order of the pion mass, the scattering amplitudes are calculated in this effective theory. For three momenta much smaller than the pion mass, it is convenient to further integrate out three momenta of the order of pion mass, which leads to the so called pionless NNEFT, and carry out the calculations in the latter. Splitting the calculation in this way we can take advantage of the modern techniques of the threshold expansions and dimensional regularization so that all integrals only depend on a single scale. There is no need to introduce a PDS scheme. The technical complexity of the N$^2$LO calculation is similar to the one in the KSW scheme, but our final expressions are simpler \cite{Soto:2009xy}. The numerical results for the phase shifts and mixing angle are also similar to the KSW ones. Hence a good description of the $^1S_0$ channel is obtained up to center of mass energies of about $35 MeV.$, but for the $^3S_1$-$^3D_1$ channel our results fail to describe data much before: for the $^3S_1$ phase shift and the mixing angle comparison with data becomes bad beyond $15 MeV.$, and for the $^3D_1$ phase shift it is never good. Particularly worrying is the fact that for the $^3S_1$ and the $^3D_1$ phase shift the N$^2$LO calculation compares worse to data than the NLO one.  The reasons of this failure can be traced back to the iteration of the OPE potential, the first N$^2$LO diagram in Fig.\ref{schannel}, which gives a very large contribution.


\acknowledgments{I thank Joan Soto for the careful reading of the manuscript. We acknowledge financial support from MICINN (Spain) grants CYT FPA FPA2007-66665-C02-01/, CPAN CSD2007-00042, the CIRIT (Catalonia) grant 2005SGR00564, and the RTNs Flavianet MRTN-CT-2006-035482 (EU). JT has been partially supported by a MICINN FPU grant ref.AP2007-01002.}


\end{document}